\documentclass[preprint,12pt]{elsarticle}
\usepackage{amssymb}
\usepackage{epsfig}
\usepackage{amsthm}
\usepackage{amsmath}
\journal{Physica A}
\begin{document}
\begin{frontmatter}

\title{Understanding dynamics of looping of a long chain polymer in solution: Wilemski-Fixman Approach}

\author{Moumita Ganguly \corref{cor1}\fnref{fn1}}
\ead{mouganguly09@gmail.com}
\address{Indian Institute of Technology Mandi, Kamand, Himachal Pradesh - 175005, India.}

\author{Aniruddha Chakraborty\corref{cor2}}
\address{Indian Institute of Technology Mandi, Kamand, Himachal Pradesh - 175005, India.}
\cortext[cor1]{Corresponding author}

\begin{abstract} 
We investigate theoretically the end-to-end looping time of a long chain polymer molecule immersed in a solvent. The dynamics of the end-to-end distance is governed by a Smoluchowski-like equation of a particle moving under the influence of a parabolic potential in presence of a Dirac delta sink of arbitrary strength and location. Using Wilemski-Fixman [ G. Wilemski and M. Fixman, J. Chem. Phys. {\bf 60}, {\it 866} (1974)] approach we calculate the looping time for a long chain molecule immersed in a solvent. We find that looping time varies with several parameters such as length of the polymer (N), bond length (b) and the relaxation time ${\tau_R}$. 
\end{abstract}


\end{frontmatter}

\section{Introduction}

\noindent
Understanding the dynamics involving loop formation of long chain polymer molecules has been an interesting research field \cite{Winnik,Haung,Lapidus,Hudgins,Wilemski,Doi,Szabo,Pastor,Portman,Sokolov,Toan}. Loop formation is understood to be an initial step in several protein \cite {Hudgins,Buscaqlia} and RNA folding \cite {Thirumalai} events. The theories of loop formation dynamics are in general very approximate \cite{Wilemski,Szabo}. In this paper, the looping time of a single polymer chain having reactive end-groups have been modeled following the work of Szabo {\it et. al.,} \cite{SzaboP} and looping time estimated using the Wilemski-Fixman approach \cite{Wilemski}. In the present model \cite{M-Physica, M-CPL} dynamics of end-to end distance is mathematically represented by a Smoluchowski-like equation for a single particle under harmonic potential in presence of a Dirac delta function sink of arbitrary strength.

\section{End-to-end motion of one-dimensional polymer}

\noindent In this section we use the simplest one dimensional description of a polymer as given by Szabo {\it et.al.,} \cite{SzaboP}. We model $2N$ monomers by a total $2N$ segments of unit length unity, where each monomer is allowed to take two possible orientations, one towards the right and the other towards the left direction. So the whole polymer can have any of $2^{2N}$ different conformations. Now we use $x$ to denote the end-to-end distance, with the value $x=2j$. After having $N$ steps the polymer segments can be either on the right $(N+j)$ or on the left $(N-j)$. So the polymer looping problem may be solved using standard theories of probability. The equilibrium end-to-end distribution $P_{0j}$ of this long polymer may be given by the following equation
\begin{equation}
P_{0j}=2^{-2N}(^{2N}_{N+j}).
\end{equation}
For an unbiased random walk problem, the probability distribution can easily be estimated by knowing the net rate of fluctuations between the polymers moving to either right or left. As the polymer molecule is immersed in a solvent, the dynamics of polymer will be determined by various intra-molecular and intermolecular forces. So the motion of polymer molecule would appear to be random. Then considering the variation of all right and left monomer segments being independent of each other the fluctuation is directed by the following rate equation.
\begin{equation}
\label{2}\frac{d}{dt}\left[^p_n\right]=\frac{1}{\tau_R}\left[
\begin{array}{cc}
-1&1\\
1& -1
\end{array}
\right]\left[^p_n\right],
\end{equation}
where the vector $[^p_n]$ represents the activity of right and left segments orientations. $\tau_R$ is the relaxation time between two configurations. Now, entire event of end-to-end looping of a polymer molecule in a solution, can be easily simplified to a simple random walk model confined in $2^{2N}$-dimensional configuration space. The individual monomer's reorientation would result in $2N$ ways to reorder a $x=2j$ conformation either to a $x=2j+2$ or $x=2j-2$ conformation and $N-J+1$ ways to reorient a $x=2j+2$ conformation into a $x=2j$ conformation. Then the resulting master equation for the end-to-end distribution $P(j,t)$ in the $(2N+1)$ -dimensional space is given by \cite{SzaboP}
\begin{equation}
\tau_R\frac{d}{dt}P(j,t)=-2NP(j,t)+(N+j+1)P(j+1,t)+(N-j+1)P(j-1,t).
\end{equation}
\noindent As we see that for long chain molecule ($N$ large), what we try to search for and measure is a distribution, {\it{i.e.,}} the probability for finding the end-to-end distance. The equilibrium distribution of  Eq.(1) may be approximated by the continuous Gaussian distribution ($x = 2 b j$)
\begin{equation}
\label{de}
P_{0}(x)=\frac{e^{-\frac{x^2}{4 b^2 N}}}{(4 \pi b^2 N)^{1/2}}
\end{equation}
Now in large $N$ limit, the end-to-end dynamics of the polymer can be represented by an equation, which is given below
\begin{equation}
\frac{\partial P(x,t)}{\partial t} = \left(\frac{4 N b^2}{\tau_R} \frac{\partial^2}{\partial x^2} + \frac{2}{\tau_R} \frac{\partial}{\partial x} x \right) P(x,t).
\end{equation}
In the above `$b$' denotes the bond length of the polymer and $x$ denotes the end-to-end distance.

\section{End-to-end reaction of one-dimensional polymer}

\noindent  Now if the two ends of the polymer molecule come sufficiently close together, a loop would be form, {\it i.e.} at $ x = a$. The occurrence of the looping reaction can be incorporated in our calculation by adding a $x$ dependent sink term $S(x)$ \cite{Swapan} (taken to be normalized {\it i.e.} $\int_{-\infty}^{\infty} S(x)dx = 1 $) in the above equation to get
\begin{equation}
\frac{\partial P(x,t)}{\partial t} = \left(\frac{4 N b^2}{\tau_R} \frac{\partial^2}{\partial x^2} + \frac{2}{\tau_R} \frac{\partial}{\partial x} x  - S(x) \right) P(x,t).
\end{equation}
The above equation can be used to describe the dynamics of loop formation. This equation is very similar to the modified Smoluchowski equation. The term $P(x,t)$ describes the probability density of end-to-end distance at time $t$. The term $S(x)$ is a function representing sink.

\section{Theory of end-to-end loop formation: Wilemski-Fixman Formalism}

\noindent Here we briefly discuss Wilemski-Fixman formalism for the end-to-end loop formation in long chains. The dynamics of a single polymer chain with reactive end-groups is modeled by Eq. (7). Wilemski and Fixman \cite{Wilemski} then derived an approximate expression for the mean first passage time from Eq. (1). In our case this mean first passage time is actually the loop closing time for the chain. The expression for this loop closing time is given by
\begin{equation}
\tau = \int^{\infty}_{0} dt \left( \frac{C(t)}{C(\infty)} - 1\right).    
\end{equation}
Where $C(t)$ is the sink-sink correlation function defined as
\begin{equation}
C(t)= \int^{\infty}_{-\infty} dx \int^{\infty}_{-\infty} dx_0 S(x) G(x,t|x_0,0)S(x_0) P_{0}(x_0),
\end{equation}
where $G(x,t|x_0,0)$ is the conditional probability that a chain with end-to-end distance $x_0$ at time $t=0$ has the end-to-end distance $x$ at time $t$ and $P_{0}(x_0)$ is the equilibrium distribution of the end-to-end distance since the chain is in equilibrium at time $t=0$. $S(x)$ is the sink function which varies with the distance of separation between chain ends. Now it is understood that the knowledge of $G(x,t|x_0,0)$ is required to calculate the sink-sink correlation function $C(t)$ and closing time $\tau$. In our model, one can show that the Green's function is given by 
\begin{equation}
G(x,t|x_0,0)= \left[\frac{1}{4 \pi N b^2 (1 - e^{- \frac{4 t}{\tau_{R}}}) }\right]^{1/2} exp\left(- \frac{\left(x - x_0 e^{ - \frac{2 t}{\tau_{R}} }\right)^2}{4 N b^2 (1 - e^{ - \frac{4 t}{\tau_{R}}})} \right)
\end{equation}
Similarly $P_{0} (x_0)$ is given by Eq. (4) of the previous section, it is nothing but the end-to-end equilibrium distribution for the long chain molecule at time $ t=0 $. Using the above expression sink-sink correlation function can be written as a double integral of position coordinates as written below
\begin{equation}
C(t)= \left[\frac{1}{(4 \pi N b^2)^2 (1 - e^{- \frac{4 t}{\tau_{R}}}) }\right]^{1/2}\int^{\infty}_{-\infty} dx S(x) \int^{\infty}_{-\infty} dx_0 S(x_0) exp\left(- \frac{\left(x - x_0 e^{ - \frac{2 t}{\tau_{R}} }\right)^2 }{4 N b^2 (1 - e^{ - \frac{4 t}{\tau_{R}}})}-\frac{x_0^2}{4 b^2 N} \right).    
\end{equation}
The above two integrals cam be evaluated analytically some specific choice of sink functions. In the next section we will assume sink to be represented by a Dirac delta function in position coordinate {\it i.e.,} $ S(x) = k_{0} \delta(x - a)$ to evaluated the above integral analytically, where 'a' is understood as capture distance.

\section{Dirac delta function sink model}

\noindent In this section we will first derive an exact analytical expression for sink-sink correlation function and then using that we will derive an expression for looping closing time. After doing analytical integration over $x$ and $x_0$ we get the following analytical expression for $C(t)$ as given below
\begin{equation}
C(t)= \frac{k_{0}^2}{(4 \pi N b^2)}\frac{1}{(1 - e^{- \frac{4 t}{\tau_{R}}})^{1/2}} exp\left(- \frac{ a^2 \left(1 - e^{ - \frac{2 t}{\tau_{R}} }\right)^2 }{4 N b^2 (1 - e^{ - \frac{4 t}{\tau_{R}}})}-\frac{a^2}{4 b^2 N} \right). 
\end{equation}
Now we take the limit, where $t \rightarrow \infty$, so that $C(t)$ becomes $C(\infty)$ and we get
\begin{equation}
C(\infty)= \frac{k_{0}^2}{4 \pi N b^2} exp\left(-\frac{a^2}{2 b^2 N} \right).    
\end{equation}
Now we can have a very simple analytical expression for the ratio $C(t)/C(\infty)$
\begin{equation}
\frac{C(t)}{C(\infty)}=\frac{1}{(1 - e^{- \frac{4 t}{\tau_{R}}})^{1/2}} exp\left(- \frac{ a^2 \left(1 - e^{ - \frac{2 t}{\tau_{R}} }\right)^2 }{4 N b^2 (1 - e^{ - \frac{4 t}{\tau_{R}}})} + \frac{a^2}{4 N b^2 }\right).     
\end{equation}
As $N$ is large, we can further simplify the above equation as
\begin{equation}
\frac{C(t)}{C(\infty)} \approx \frac{1}{(1 - e^{- \frac{4 t}{\tau_{R}}})^{1/2}}\left( 1 - \frac{ a^2 \left(1 - e^{ - \frac{2 t}{\tau_{R}} }\right)^2 }{4 N b^2 (1 - e^{ - \frac{4 t}{\tau_{R}}})} + \frac{a^2}{4 N b^2}\right).   
\end{equation}
We can re-write the above equation as
\begin{equation}
\frac{C(t)}{C(\infty)} \approx \frac{1}{(1 - e^{- \frac{4 t}{\tau_{R}}})^{1/2}}\left( 1 - \frac{ a^2 \left(1 + e^{ - \frac{4 t}{\tau_{R}}} - 2 e^{ - \frac{2 t}{\tau_{R}} } \right)}{4 N b^2 (1 - e^{ - \frac{4 t}{\tau_{R}}})} + \frac{a^2}{4 N b^2 }\right). 
\end{equation}
Now we assume the case where the case where $\tau_{R}$ is small, so that  $(1 \pm e^{-\frac{4 t}{\tau_{R}}})$ approach $1$. So Eq. (15) becomes
\begin{equation}
\frac{C(t)}{C(\infty)} \approx \left( 1 - \frac{ a^2 \left(1  - 2 e^{ - \frac{2 t}{\tau_{R}} } \right)}{4 N b^2} + \frac{a^2}{4 N b^2 }\right).
\end{equation}
After simplification, we get
\begin{equation}
\frac{C(t)}{C(\infty)} = 1  + \frac{e^{-\frac{2 t}{\tau_{R}}}}{2 N b^2}.
\end{equation}
So the loop closing time $\tau$, may be calculated by doing the following integration
\begin{equation}
\tau = \frac{1}{{2 N b^2}} \int^{\infty}_{0} dt e^{-\frac{2 t}{\tau_{R}}} = \frac{\tau_{R}}{4 N b^2}.
\end{equation}
So loop closing time is directly proportional to the relaxation time $\tau_{R}$ and inversely proportional to the length of the polymer $N$ and it is inversely proportional to the square of the bond length $b$.

\section{Conclusions:}
\noindent Here we have derived an analytical expression for loop closing time for long chain polymer molecule in solution. Our result shows how looping time varies with several parameters such as length of the polymer ($N$), bond length ($b$) and the relaxation time (${\tau_R}$). We have found that loop closing time decreases with increase in bond length, so a polymer with longer bond will form loop quickly - it is actually the most sensitive parameter and longer polymer will form loop very quickly. Polymer with larger relaxation time will form loop quickly.   

\section{Acknowledgments:}
\noindent One of the author (M.G.) would like to thank IIT Mandi for HTRA fellowship and the other author thanks IIT mandi for providing CPDA grant.

\end{document}